\documentclass[a4paper,11pt]{article}

\usepackage{lineno}
\usepackage{xcolor}
\usepackage{booktabs}
\usepackage{longtable}
\usepackage{algpseudocode}
\usepackage[normalem]{ulem}
\usepackage{slashed}
\usepackage{graphicx}
\usepackage{subcaption}
\graphicspath{ {./figures/} }

\usepackage{jheppub} 

\usepackage{amsmath}
\usepackage{amsthm}
\usepackage{mathtools}
\usepackage{amssymb}
\usepackage{physics}

\title{Effective two-loop background contributions to $g_e-2$}

\author[a,b]{Jason L. Evans}
\author[a,b]{and Junyuan Lyu}

\affiliation[a]{Tsung-Dao Lee Institute, Shanghai Jiao Tong University, Shanghai 201210, China}
\affiliation[b]{School of Physics and Astronomy, Shanghai Jiao Tong University, Shanghai 200240, China}

\emailAdd{lvjunyuan@sjtu.edu.cn}
\emailAdd{jlevans@sjtu.edu.cn}

\abstract{Recent work has shown the utility of examining background effects from ultralight dark matter on precision measurements. This effect enhances seemingly benign contributions to the level that experiments are sensitive to them. In this work, we examine the consequences of the effective axion-like particles coupling to photons in a background of axion-like particle dark matter. This analysis leads to new constraints on the product of the axion electron and axion photon coupling. Furthermore, since the axion photon coupling is generated, at one loop, by a pseudoscalar Yukawa coupling, this calculation can also be applied to constraining the axion-like particles coupling to the electron at the two-loop level. This contribution often dominates due to the fact that the one-loop contribution to the anomalous magnetic moment of the electron from a pseudoscalar is momentum suppressed.  }

\makeatletter\def\@fpheader{~}\makeatother

\begin{document}
\maketitle
\flushbottom

\section{Introduction}
The composition of dark matter is still an unsolved problem in particle physics and cosmology. Currently, the allowed mass range for dark matter is extremely large and the nature of its interactions are also not well understood. 

Phenomenologically, ultralight dark matter is a promising candidate for dark matter \cite{Ferreira_2021}. Its rather long De Broglie wavelength has astrophysical implications. It sets a lower bound on the size of dark matter density fluctuations in the universe. This lower bound on the size of dark matter fluctuations can potentially solve the missing satellite problem \cite{Klypin_1999, Ferreira_2021, Witten_2017, Lee_2010}, since it would exclude the possibility of smaller galaxies forming. Another problem this long wavelength dark matter can potentially solve is the core-cusp problem \cite{Blok_2010, Schive_2014, Ferreira_2021, Witten_2017}. Since density fluctuations must be larger than the De Broglie wavelength, the galaxies must be cored\footnote{Some recent measurements of the Lyman-$\alpha$ forest \cite{Iršič_2017, Rogers_2021} call in to question this advantage of ultralight dark matter. However, these measurements are sensitive to the uncertainties like those in the thermal state of the intergalactic medium \cite{Jackson_2024}.} .

One particularly promising ultralight dark matter candidate, sometimes referred to as weakly interacting slim particles (WISPs) \cite{generic_2020}, are the QCD axion \cite{PQ_1977, Kim_2010, Luca_2020} and axion-like particles \cite{Witten_2006, Arvanitaki_2010, Ringwald_2014, Baue_2017, Irastorza_2018, Cornella_2020, Buen_2021, Bauer_2022}. The QCD axion, in particular, stands out among other dark matter candidates as it is motivated by the strong CP problem \cite{Kim_2010}. Axion like particles have their own motivations stemming from string theory \cite{Witten_2006, Arvanitaki_2010,Ringwald_2014}.

Production of these pseudoscalar ultralight dark matter candidates are accomplished through a misalignment mechanism \cite{Luca_2020, generic_2020, Marsh_2016}. This mechanism generates an oscillating axion vacuum expectation value (vev) that behaves as a cold fluid. The gravitational interactions that form galaxies then decoherence the axion leaving a quasi-coherent state of dark matter \cite{Ferreira_2021, Allali_2020}.

Because axion and axion-like particle dark matter is very light and very weakly interacting, detection is quite challenging.  Since the inception of the invisible axion 40 years ago, detection has been a key challenge for these models.  A great deal of progress has been made on detection, but experiments have struggled to reach the accuracy needed to probe the QCD axion. 

Although axion-like particles  fill out a much broader parameter space, much of its parameter space remains unexplored. The current best constraints on the axion-like particle couplings we wish to examine,  the axion-photon interaction $g _{a \gamma }$ and axion-electron interaction $g _{ae}$, come from astrophysics and cosmology.  The strongest constraint on $g _{a \gamma }$ are from the CERN Axion Solar Telescope (CAST) \cite{CAST_2017} which searches for x-rays generated from the processes $a \to \gamma \gamma $ and has led to the constraints $g _{a \gamma }<6.6\times 10 ^{-11}\, \mathrm{GeV}^{-1}$ for axion-like particle mass $m _{a}\lesssim 0.02\, \mathrm{eV}$. The recent spectroscopy of the cluster-hosted quasar H1821+643 \cite{H1821+643_2021} resulted in the exclusion $g _{a \gamma }>6.3\times 10 ^{-13}\, \mathrm{GeV}^{-1}$ for $m _{a}\lesssim 10 ^{-12}\, \mathrm{eV}$. For the axion-electron interaction $g _{ae}$, the electron recoil experiments of XENON1T \cite{XENON1T_2020} and XENONnT \cite{XENONnT_2020} have excluded $g _{a e}>2\times 10 ^{-12}$ for ALP mass $m _{a}\lesssim 100\, \mathrm{eV}$. Red-giant observations \cite{Red-giant_2020} have pushed this bound to $g _{a e}< 1.3\times 10 ^{-13}$.

Since ALPs are generally quite light, the dark matter number density is enormous. This leads to a background with a large occupation number. In this case, the ground state is no longer the vacuum, $| 0\rangle $, but instead a bosonic state $| n (k) \rangle $, where $n (k)$ is the occupation number of the background dark matter. Propagation of the dark matter in this background is now between bosonic states and is modified to 
\begin{equation}\label{1.1}
    \Delta_{F}(q)\equiv\mathcal{F} \left\{ \left\langle n _{a}|\mathcal{T}  \phi (x) \phi (y) |n _{a} \right\rangle\right\}  =\frac{i}{q^{2}-m_{a}^{2}+i\epsilon }-2\pi n_{a} (q)\delta(q^{2}-m_{a}^{2})~,
\end{equation}
for a scalar boson, where $\mathcal{F}$ is the Fourier transformation.  For a thermal distribution, Eq. (\ref{1.1}) is nothing but the propagator of real-time formalism in thermal field theory.  Since the propagator in Eq. (\ref{1.1}) can be derived solely based on the properties of a boson and not on the shape of $n(k)$, this propagator can be applied to any background of bosons.  

Here we will examine the effect of this background enhanced propagation for axion-like particle dark matter. We will focus on the axion (effective) coupling to photons. Although this couplings is effectively two-loops, we will still be able to place new constraints on $g_{a\gamma}$ and $g_{ae}$.    

This article is structured as follows: In Sec. \ref{2}, we introduce the effective couplings of ALPs with electrons and photons as well as the background formalism. Next, we calculate effective two-loop background diagrams. In Sec. \ref{gauge}, we justify our result by showing that even in  presence of a dark matter background gauge-invariance is not broken and electric charge is not renormalized. In Sec. \ref{anal}, we extract the effect Hamiltonian from the vertex correction then give the analytical expression for the electron $g-2$ in the presence of an axion dark matter background. Lastly, we present our constraints of $g_{ae}$ and $g_{a\gamma}$ which are shown to be stronger than previous constraints.  We conclude in Sec. \ref{5}. Technical details of our calculations are relegated to the two appendices.

\section{Background Effect on One-Loop Diagrams}\label{2}
Quantum electrodynamics (QED) gives a very precise prediction for the electron magnetic momentum. According to recent experimental results \cite{ Fan_2022}, 
\begin{equation}
\delta\left(\frac{g_e}{2}\right)=\frac{g_e}{2}(\text {Measured})-\frac{g_e}{2}(\text {Theory})=(3.41 \pm 1.64) \times 10^{-13} 
\end{equation}
Because of this precise agreement of theory and experiment, we are justified in considering that axion-like dark matter background corrections to $g-2$ are the source of the dominant deviation from the SM prediction. This allows us to use the experimental measurements of the electron magnetic moment to constrain axion-like particles. 

\subsection{The effective Lagrangian}\label{eff}
In this work, we consider the following axion-like dark matter couplings
\begin{equation}\label{original}
    \mathcal{L}_{\mathrm{eff}}\supset \frac{\bar g_{ae}}{2m_{e}} \partial_{\mu }a \bar{\psi} \gamma^\mu \gamma_5 \psi+\frac{1}{4}g_{a\gamma }a F \tilde{F}+g_{ae}a\bar\psi i \gamma_5 \psi
\end{equation}
where $a$ is a light pseudoscalar boson which we refer to as an axion-like-particle (ALP). In this paper, we discuss both the Yukawa like coupling and derivative coupling.

\subsection{Background formalism}\label{background}
In models with an axion like particle dark matter background, the propagator for the axion like particle becomes 
\begin{equation}
    \Delta_{F}(q)=\frac{i}{q^{2}-m_{a}^{2}+i\epsilon }-2\pi n_{a} (q)\delta(q^{2}-m_{a}^{2})
\end{equation}
where $n_{a}(q)$ is some unknown occupancy number of a background axion like dark matter. We will estimate this later (see Sec. \ref{anal}).

Before calculating the background contribution, we mention the vertex correction from virtual ALPs to assure the reader it is subdominant to the background contribution. The contribution of virtual particles 
gives a contribution of approximately \cite{Cornella_2020, Buen_2021, Bauer_2022}

\begin{equation}\label{virtual}
    \Delta a_e \approx-\frac{g _{ae}^{2}}{16 \pi^2}-\frac{m _{e}g _{ae} g _{a \gamma}}{8 \pi^2} 
\end{equation}\label{virtual2}
As we will see below, the background contribution has the same coupling suppression but is accompanied by a factor of order $\frac{\rho _{a}}{m _{a}^{2}m _{e}^{2}}$. This factor leads to a drastic enhancement in the case of ultralight dark matter. 

We now calculate the background contribution. We begin with the electron self energy, which we break up into two contributions 
\begin{equation}
    \Sigma(k)=\Sigma_{0}(k)+\Sigma_{\mathrm{BG} }(k)~,
\end{equation}
where $\Sigma_{0}(k)$ is the contribution from virtual particles and $\Sigma_{\mathrm{BG} }(k)$ is background contribution. Generally speaking, $\Sigma_{\mathrm{BG} }(k)$ can be decomposed into
\begin{equation}
    \Sigma_{\mathrm{BG} }(k)=B(k)+\left(\slashed{k}-m_{e}\right)C(k)+\slashed{D}(k)~,
\end{equation}
where the form of $B (k)$, $C (k)$ and $D_{\mu }(k)$ depend on the electron interaction with the background. For a pseudoscalar Yukawa interaction, these functions are
\begin{equation}
    \begin{aligned}
        B(k)&=0\\
        C(k)&=g_{ae}^{2}\int \dfrac{d ^{4} \Xi _{q}}{(2\pi)^{3}}\dfrac{1}{(k+q)^{2}-m_{e}^{2}}\\  
        D_{\mu }(k)&=g_{ae}^{2}\int \dfrac{d ^{4} \Xi _{q}}{(2\pi)^{3}}\dfrac{q _{\mu }}{(k+q)^{2}-m_{e}^{2}}~,\\ 
    \end{aligned}
\end{equation}
where
\begin{equation}
    d^4 \Xi  _q=d^4 q \bar{n}(q) \delta(q^2-m_a^2)~.
\end{equation}

\begin{figure}
    \centering
    \includegraphics{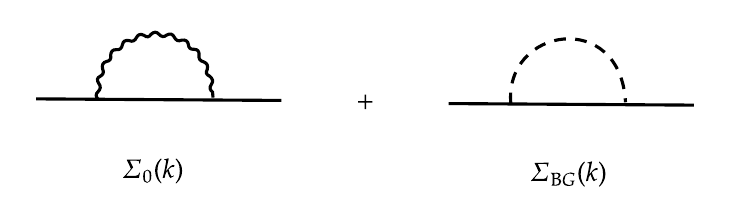}
    \hfill
    \caption{Feynman diagrams of the electron self-energy. The left diagram is the vacuum contribution from virtual photons. The right diagram is the background contribution. }
    \label{2-self-energy.pdf}
\end{figure}

The non-standard contribution to the wave function renormalization from $D^\mu(k)$ is a result of the background breaking Lorentz invariance. To cope with this non-standard form of the wave function renormalization, we introduce background dependent spinors and Lorentz violating mass counter terms\footnote{The validity of these background dependent spinors are born out in the proof of charge non-renormalization and gauge invariances  as seen in \cite{Donoghue_1985} and later here in Sec. \ref{gauge}.} \cite{Donoghue_1985}, which satisfy

\begin{equation}
    \left[\slashed{k}-m_{e}+\Sigma_{\mathrm{BG} }(k)\right]u_{n}(k)=0~.
\end{equation}
The dispersion relations are modified by the background to give
\begin{equation}
    \tilde{\slashed{k}}-\tilde{m}\equiv \slashed{k}-m_{e}+\Sigma_{\mathrm{BG} }(k)~.
\end{equation}
The spinors then satisfy the following relationship
\begin{equation}
    \begin{aligned}
        u_{n}(k)^{\dagger}u_{n}(k)&=1 \\ 
        \sum_{\text {spin }} u_n(k) \bar{u}_n(k)&=\frac{\tilde{\slashed{k}}+\hat{m}}{2 \tilde{E}}~,
    \end{aligned}
\end{equation}
where
\begin{equation}
    \tilde{E}=\tilde{p}^{0}=E\left(1-\left(C+\dfrac{D^{0}}{E}\right)\right)~.
\end{equation}
We can then calculate the propagator $\langle \psi \bar \psi\rangle$ which gives
\begin{equation}
    S^R_{F}(x-y)=-\int \frac{d^3 k}{(2 \pi)^3}\left[\Theta(x_0-y_0) \frac{\tilde{\slashed{k}}+\tilde{m}}{2 \tilde{E}} e^{-i k \cdot(x-y)}-\Theta(y_0-x_0) \frac{\tilde{\slashed{k}}-\tilde{m}}{2 \tilde{E}} e^{i k \cdot(x-y)}\right]~.
\end{equation}
We can also get the propagator from resuming the 1-PI contributions involving $\Sigma(k)$ to get 
\begin{equation}
    S^R(x-y)  =i \int \frac{d^4 k}{(2 \pi)^4} \frac{Z_2^{-1} e^{-i k \cdot(x-y)}}{\slashed{k}-m_{\mathrm{R}}+\Sigma_{\mathrm{BG} }(k)+i \epsilon }~.
\end{equation}
Comparing these two contributions, we can determine the wave-function renormalization  
\begin{equation}
    \begin{aligned}
        Z_{2}^{-1}&=\left(1+C+\dfrac{m}{E}\dfrac{d}{dE}\left(\dfrac{k\cdot D}{m_{e}}\right)-\dfrac{D^{0}}{E}\right)\\
        \delta m_{e}&=-\frac{k \cdot D(k)}{m_{e}}\\ 
        E&=\sqrt{\vec{k}^{2}+m_{\beta}^{2}}~,
    \end{aligned}
\end{equation}
where $m_{\beta}\equiv m_{e}+ \delta m_{e}$ is the physical electron mass in the background.

From the previous discussion, we see that the new self-energy $\Sigma_{\mathrm{BG} }$ shifts the pole mass and gives a corrected wavefunction renormalization counter term. Apart from the Lorentz violating contribution to the wave function renormalization, which can be accounted for by a counter term, the fermions are still effectively a free Dirac field. 

\subsection{The One-Loop Contribution}
\begin{figure}
    \centering
    \includegraphics[width=14cm]{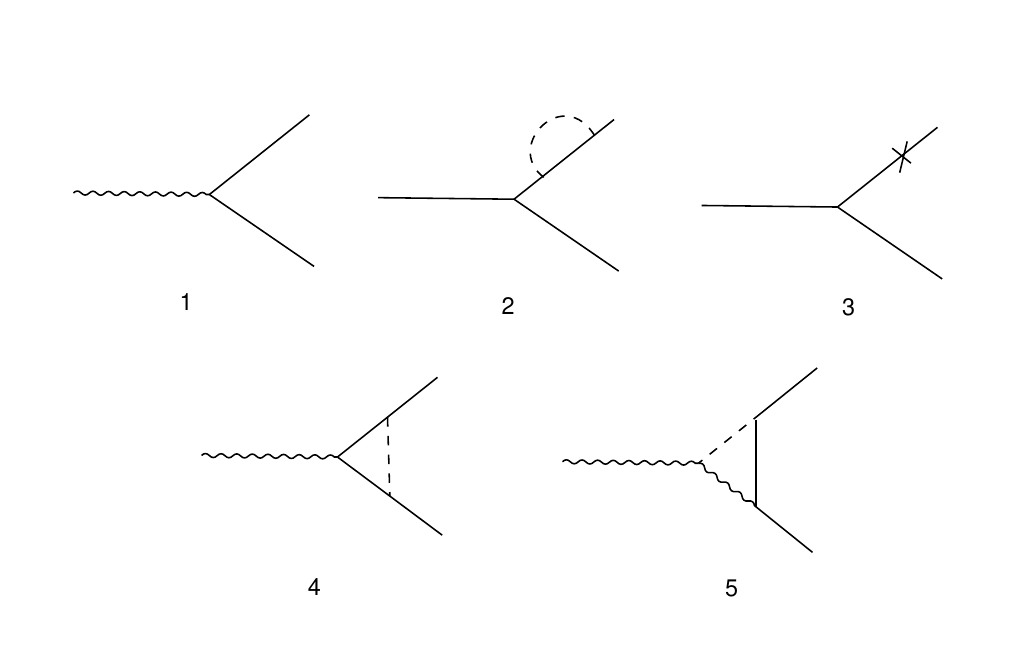}
    \caption{Feynman diagrams for the background contributions to electron $g-2$. 
    }
    \label{vertex1.pdf}
\end{figure}
We consider all one-loop background contributions to the electron magnetic moment from axion like particles. There are five total diagrams.  
The first three diagrams found in Fig. \ref{vertex1.pdf} give
\begin{equation}\label{123}
    \begin{aligned}
        i\mathcal{M}^{\mathrm{\uppercase\expandafter{\romannumeral1}}}_{\mu }&\equiv  i\mathcal{M}^{\mathrm{Tree}}_{\mu }+i\mathcal{M}^{\mathrm{SE}}_{\mu }+i\mathcal{M}^{\mathrm{CT}}_{\mu }\\   
        &=(-ie)\bar{u}_{n}(\bar{k}) \gamma_{\mu }\left[1-\frac{1}{2}C(k)-\frac{1}{2}\dfrac{m}{E}\dfrac{d}{dE}\left(\dfrac{k\cdot D(k)}{m}\right)+\frac{1}{2}\dfrac{D_{0}(k)}{E}+\left(k\leftrightarrow \bar{k}\right)\right]u_{n}(k)~.
    \end{aligned}
\end{equation}
The last two diagrams in Fig. \ref{vertex1.pdf} give\footnote{We have kept only leading order in $\Delta k$ and $m_{a}^{2}$ .}
\begin{equation}
    \begin{aligned}
        i\mathcal{M}^{\mathrm{\uppercase\expandafter{\romannumeral2}}}_{\mu }&=i\mathcal{M}^{4}_{\mu }+i\mathcal{M}^{5}_{\mu }\\ 
        &=(-ie)\bar{u}_{n}(\bar{k})\left[\frac{1}{2}C(k)+\frac{1}{2}\frac{d }{d k_\mu}\left(\dfrac{k\cdot D(k)}{m_{e}}\right)-\frac{1}{2}\dfrac{D_{\mu }(k)}{m_{e}}+\left(k\leftrightarrow \bar{k}\right)+F_{\mu }\left(k,\Delta k\right)\right] u_{n}(k)~,
    \end{aligned}
\end{equation}
where $F_{\mu }\left(k,\Delta k\right)$ comes from diagram 4 and diagram 5 in Fig. \ref{vertex1.pdf} and is defined in Appendix \ref{process}. This contribution cannot be written in terms of the  generic functions $C(k)$, $D_\mu (k)$, and  $B(k)$.  For axion dark matter and the couplings $g_{ae}$ and $g_{a\gamma}$, $F_{\mu }\left(k,\Delta k\right)$ is
\begin{equation}\label{FF}
    F_{\mu }\left(k,\Delta k\right)=\dfrac{\Delta k^{\alpha }}{2m_{e}}\left[i\sigma_{\nu  \alpha }\cdot \left(\dfrac{1}{2}R \slashed{\bar{I}}(k)\delta^{\nu }_{\mu }-\frac{1}{4}\bar{I}_\mu^\nu+\frac{1}{2}R \frac{k_\mu}{m_e} \bar{I}^\nu(k)-2I^{\nu }_{\mu}+2 R I_{B}\delta^{\nu }_{\mu }\right)+\dfrac{1}{2}R\bar{I}_{\alpha }(k)\right]~,
\end{equation}
where $R\equiv  {m_{a}^{2}}/{m_{e}^{2}}$ and  
\begin{equation}\label{int}
    \begin{aligned}  
        I_{A}(k)&\equiv g_{ae}^{2}\int  \dfrac{d^{4}\Xi_{q}}{(2\pi)^{3}}\dfrac{m^{2}_{e}}{(q\cdot k)^{2}}\\ 
        I_{B}(k)&\equiv g_{ae}g_{a\gamma }\int  \dfrac{d^{4}\Xi_{q}}{(2\pi)^{3}}\dfrac{m^{3}_{e}}{2(q\cdot k)^{2}}\\ 
        I_{\mu\nu }(k)&\equiv  g_{ae}g_{a\gamma }\int  \dfrac{d^{4}\Xi_{q}}{(2\pi)^{3}}\dfrac{m _{e}q_{\mu }q_{\nu }}{2(q\cdot k)^{2}}\\
        \bar{I}_{\mu\nu }(k)&\equiv g_{ae}^{2}\int  \dfrac{d^{4}\Xi_{q}}{(2\pi)^{3}}\dfrac{q_{\mu }q_{\nu }}{(q\cdot k)^{2}}\\
        \bar{I}_{\mu }(k)&\equiv g_{ae}^{2}\int \dfrac{d ^{4}\Xi_{q}}{(2 \pi)^3}\dfrac{m_{e}^{3}q_{\mu }}{\left(k \cdot q\right)^{3}}~.\\
    \end{aligned}
\end{equation}
The total vertex from these one-loop diagrams is
\begin{equation}\label{tot}
    \begin{aligned}
        i\mathcal{M}^{\mathrm{TOT}}_{\mu }&=i\mathcal{M}^{\mathrm{\uppercase\expandafter{\romannumeral1}}}_{\mu }+i\mathcal{M}^{\mathrm{\uppercase\expandafter{\romannumeral2}}}_{\mu }\\   
        &=(-ie)\bar{u}_{n}(\bar{k})\left[\gamma_{\mu }+G_{\mu }(k)+G_{\mu }(\bar{k})+F_{\mu }\left(k,\Delta k\right)\right]u_{n}(k)~,
    \end{aligned}
\end{equation}
where
\begin{equation}
    G_{\mu }(k)\equiv  \frac{1}{2}\left(\dfrac{d}{d k^\mu}-\gamma_{\mu }\dfrac{m}{E} \frac{d}{d E}\right)\left(\frac{k \cdot D(k)}{m_e}\right)+\gamma_{\mu }\frac{D_{0}(k)}{2E}-\frac{D_{\mu }(k)}{2m_{e}}~.
\end{equation}
More details describing how to derive the above contribution can be found in Appendix \ref{process}. Next, we perform some self-consistency checks to assure the read of our results.

\section{Charge Non-renormalization and Gauge Invariances}\label{gauge}
Although introducing a background field breaks Lorentz invariance, the electric charge must be unchanged and gauge invariance must be preserved. If charge is unrenormalized and gauge invariance is preserved, we can be confident the other physical effects are legitimate (Similar questions are considered in \cite{Donoghue_1985, Jason_2022, Jason_2023}).

\subsection{Charge Non-renormalization}
In the vacuum, non-renormalization of the charge can be seen in the expression 
\begin{equation}\label{charge-non}
     \mathcal{M}^{\mathrm{VAC}}_{0 }(k,k)=-\bar{u}(k)e_{\mathrm{R}} \gamma _{0}u(k)=-e_{\mathrm{R}}~.
\end{equation}
This relationship should be unaffected by the background. 
To show that Eq. (\ref{charge-non}) is preserved in our calculation, we set $\Delta k=0$ in Eq. (\ref{FF}) to get
\begin{equation}
    \begin{aligned}
        \mathcal{M}^{\mathrm{TOT}}_{0 }(k,k)&=-e_{\mathrm{R}}\bar{u}_{n}(k)\left[\gamma_{0}+\left(-\frac{m_{e}}{E}\gamma_{0}+1\right)\left(\frac{d}{d E}\left(\frac{k \cdot D(k)}{m_e}\right)-\frac{D_{0}}{m_{e}}\right)\right]u_{n}(k)\\ 
        &=-e_{\mathrm{R}}\bar{u}_{n}(k)\gamma_{0}u_{n}(k)=-e_{\mathrm{R}}~.
    \end{aligned}
\end{equation}
In the last line, we have used Gordon decomposition and find that charge is indeed not renormalized by the axion-like particle background. 

\subsection{Gauge invariances}
Next, we show that the background has not broken gauge invariance. The retention of gauge invariance is shown by checking that the Ward identity, 
\begin{equation}\label{gauge2}
    \Delta k ^{\mu }\mathcal{M}^{\mathrm{TOT}}_{\mu }=0~,
\end{equation}
is unaffected. The contribution $\Delta k^\mu F_\mu(\Delta k,k)$ vanishes. 
The following contributions remain, 
\begin{equation}
    \begin{aligned}
        \Delta k ^{\mu }\mathcal{M}^{\mathrm{TOT}}_{\mu }&=-e\bar{u}_{n}(\bar{k})\gamma_{\nu }\left[D^{\nu }(k)-D^{\nu }(\bar{k})+\Delta k^{\mu }\left(\frac{1}{2}\frac{d D^{\nu }(k)}{d k^{\mu }}+\frac{1}{2}\frac{d D^{\nu }(\bar{k})}{d \bar{k}^{\mu }}\right)\right]u_{n}(k)\\ 
        &=\mathcal{O}\left((\Delta k)^{3}\right)=0~.\label{eq:WardIdent}
    \end{aligned}
\end{equation}
The expression above uses 
\begin{equation}
    \bar{u}_{n}(\bar{k})\left(\gamma_{\mu }\Delta k^{\mu }\right)u_{n}(k)=\bar{u}_{n}(\bar{k})\gamma_{\mu }\left[D^{\mu }(k)-D^{\mu }(\bar{k})\right]u_{n}(k)~,
\end{equation}
which is required by background dependent spinors. Expanding Eq. (\ref{eq:WardIdent}) in $\Delta k_\mu$, we see that the leading order contribution is of order $(\Delta k)^3$, which is higher order in $\Delta k$ than we have considered. Thus, our calculation is consistent with our intuition about the background. Now that we have gained confidence in our calculation, we will next examine the effect this has on the experimentally measured observables.

\section{Analytical Results}\label{anal}
Now, we use the experimental results for the measurement of $g-2$ to constrain the loop contribution to the photon vertex.  The effect on experimental results can be found by including this correction to the Dirac equation and solving for the Hamiltonian. This will allows us to peel of the correction to the magnetic moment from the background fields and compare the resulting magnetic moment to the experimental results. The consistency of the standard model prediction and measurement will allow us to constrain the couplings $g_{ae}$ and $g_{a\gamma }$.

\subsection{The effective Hamiltonian}
To get the effective Hamiltonian, we use the Foldy-Wouthuysen transformation as was done in \cite{Donoghue_1985}. This transformation corresponds to
\begin{equation}
    \mathcal{H} =\exp \left[-i \frac{1}{2} \phi \rho_2\right] H \exp \left[i \frac{1}{2} \phi \rho_2\right]~,
\end{equation}
where $\rho_2=i\gamma_0\gamma_5$ with more discussion in Appendix \ref{process} and 
\begin{equation}\label{tan}
    \tan \phi  =\dfrac{\vec{\sigma}\cdot \vec{\pi}}{m_{e}}~.
\end{equation}
After some algebra, we find \footnote{The details can be found in Appendix \ref{Ham}.}
\begin{equation}\label{effH}
    \mathcal{H}=E-\frac{e}{2E}\left(\vec{L}\cdot \vec{B}\right)\left(1+\frac{D^0(k)}{E}\right)-\frac{e}{2E}\left(\vec{\sigma}\cdot \vec{B}\right)\left(1+\frac{D^0(k)}{E_k}+\mathcal{W}(k) \right)~,
\end{equation}
where $\mathcal{W}(k) $ is the contribution due to the background and is defined as
\begin{equation}
    \mathcal{W}(k)=-\frac{E_{k}}{m_{e}}\left(\frac{1}{2}R I_A(k)-2RI_{B}(k)\right) +\frac{1}{2}R\bar{I}^{0}+\frac{|\vec{k}|^{2}}{4m_{e}^{2}}R\bar{I}^0(k)~.
\end{equation}

\subsection{New Constraints}
From the Hamiltonian in Eq. (\ref{effH}), we see that both the spin and cyclotron frequencies are modified by the background. Thus, they can be constrained by the experimental data \cite{Fan_2023, Fan_2022} on the electron $g-2$. 

The cyclotron and spin frequency can be read off from Eq. (\ref{effH}), which gives 
\begin{equation}
    \begin{aligned}
        \omega_{c}&= \frac{eB}{E}\left(1+\frac{D^0(k)}{E}\right)   \\  
        \omega_{s}&=\omega_{c}\left(1+\frac{\alpha}{2\pi}\frac{E}{m_{e}}+\mathcal{W}(k)\right)~,
    \end{aligned}
\end{equation}
where we have added the one-loop QED contribution to the electron magnetic moment with zero background. $\mathcal{W}(k)$ is the predicted deviation from the SM in the measurement if ultralight axion like dark matter is present.

In order to numerically compare our prediction with experiment, we need to estimate the dark matter occupation number. Because the dark matter velocity is very non-relativistic \cite{Foster_2018}, the occupation number should be similar to that of a Bose-Einstein condensate state,  
\begin{equation}\label{distribution}
    n_{a}(q)=\frac{\rho_{a}}{m_{a}}(2\pi)^{3}\delta ^{3}(\vec{q})~,
\end{equation}
where $\rho_{a}\approx 0.35\, \mathrm{GeV}/\mathrm{cm}^{3}$ \cite{Kafle_2014} is the approximate dark matter density near earth and $m_a$ is the dark matter mass. Using this simple expression for the occupation number, the integrals in Eq. (\ref{int}) can be evaluated to get
\begin{equation}\label{www}
    \mathcal{W}(k)=-\frac{\kappa }{4}\frac{|\vec{k}|^{2}}{E^{2}}+2\zeta ~,
\end{equation}
where 
\begin{equation}
        \kappa\equiv \frac{g_{ae}^{2}\rho_{a}}{m_{a}^{2}m_{e}E}
\end{equation}
\begin{equation}
    \zeta \equiv \frac{g_{ae} g _{a\gamma }\rho_{a}}{2m_{a}^{2}E}~.
\end{equation}

We can determine the correction to the measured quantity, $R_f$ by expanding it in $\delta\omega_a$ and $\delta \omega_c$. This gives us the following relationship,  
\begin{equation}
R_f=\frac{\omega_a}{\omega_c}=\frac{\omega_{s }-\omega_c}{\omega_c} \simeq R_{f_0}\left[1+\frac{\delta \omega_a}{\omega_{a_0}}-\frac{\delta \omega_c}{\omega_{c_0}}\right]~,
\end{equation}
where $\delta \omega _{a,c}$ are background corrections to the frequencies, $R_{f_0}$ and $\omega_{(s,c)_0}$ are SM predicted values. This gives us
\begin{equation}
\frac{\delta R_f}{R_{f_0}}\simeq \frac{\delta \omega_a}{\omega_{a_0}}-\frac{\delta \omega_c}{\omega_{c_0}}\simeq\frac{\delta \omega_a}{\omega_{a_0}}\simeq \frac{2\pi}{\alpha}\frac{m}{E} \left[ -\frac{\kappa }{4}\frac{|\vec{k}|^{2}}{E^{2}}+2\zeta\right]~.
\end{equation}

Next, we propagate the experimental error to determine the constraint on $R_f$'s from the error in $\omega_a$ and $\omega_c$.  This gives 
\begin{equation}
\frac{\Delta R_f}{R_{f_0}}=\frac{\Delta \omega_a}{\omega_{a_0}}-\frac{\Delta \omega_c}{\omega_{c_0}}~,
\end{equation}
where $\Delta \omega_{c,a}$ are the error in the measurement of these quantities. The error for these quantities can be found in \cite{Fan_2022} and is
\begin{equation}
\frac{\Delta \omega_c}{\omega_c} \simeq  2 \times 10^{-11}, \quad \frac{\Delta \omega_a}{\omega_a}\simeq 4 \times 10^{-12}~.
\end{equation}
This gives the following constraint 
\begin{equation}
\frac{\delta \omega_a}{\omega_{a_0}}\simeq \frac{2\pi}{\alpha}\frac{m_e}{E} \left\lvert  -\frac{\kappa }{4}\frac{|\vec{k}|^{2}}{E^{2}}+2\zeta\right\rvert< \frac{\Delta \omega_a}{\omega_a}=4 \times 10^{-12}~.
\end{equation}

The cutting edge experimental measurement of the anomalous magnetic moment of the electron uses an ultra non-relativistic particle with $|\vec{k}| \approx  10 ^{-4}\, \mathrm{MeV}$ \cite{Fan_2023}. This leads to a drastic suppression to the correction to $g-2$ from the contribution proportional to $g_{ae}^2$. 

\begin{figure*}
    \begin{tabular}{cc}
    \\
    \includegraphics[width=0.48\textwidth]{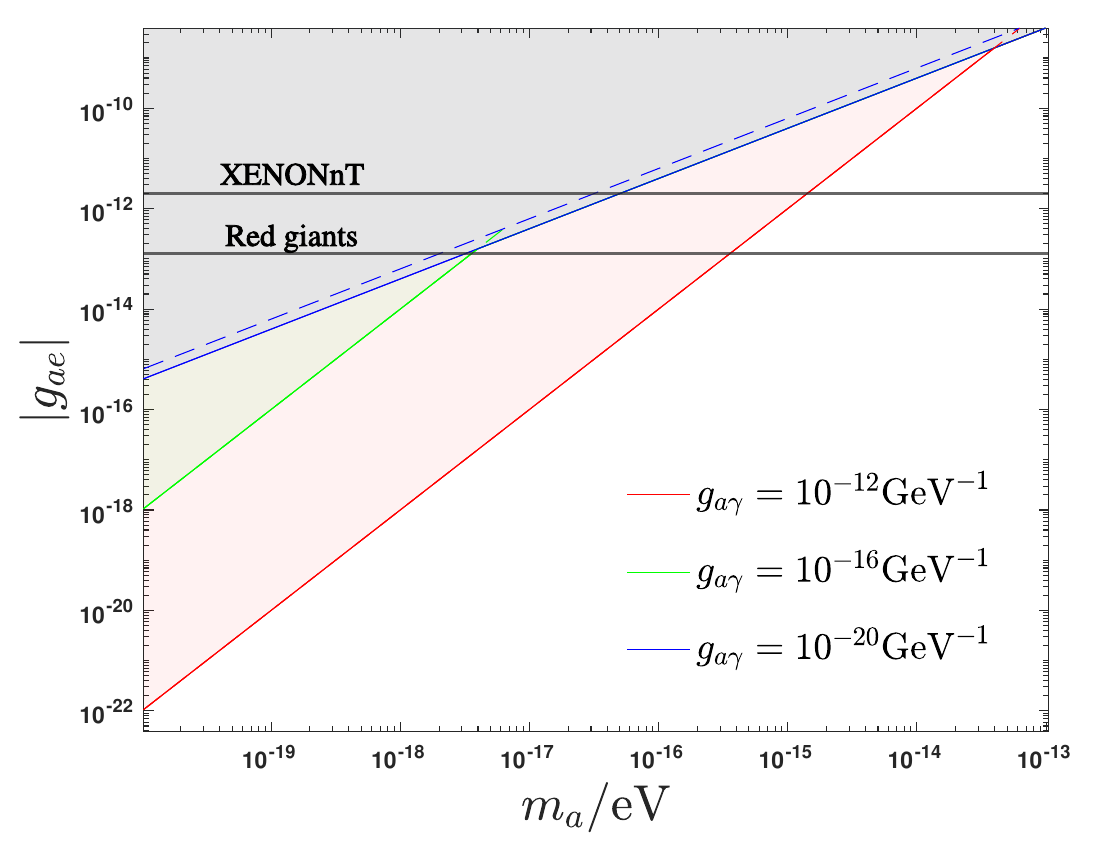}&
    \includegraphics[width=0.48\textwidth]{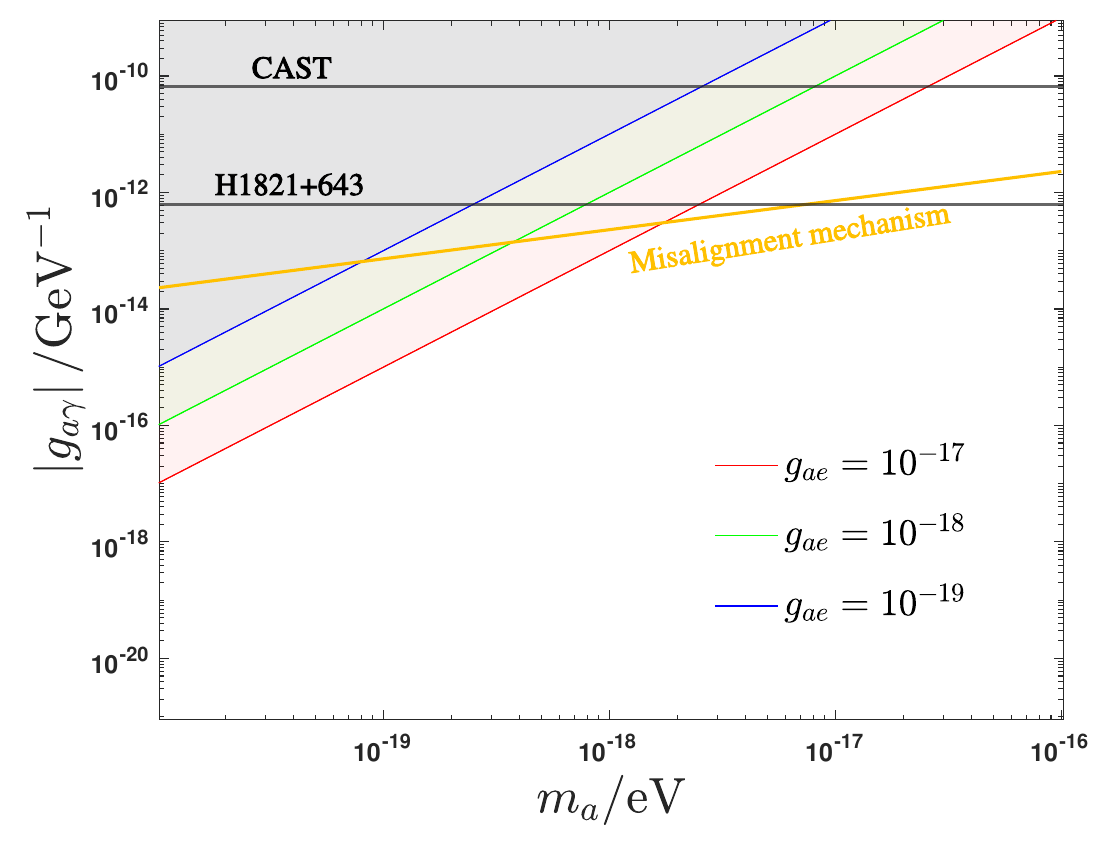}\\

    \end{tabular}
    \caption{Constraints on the axion-electron coupling (left) and axion-photon coupling (right). For derivative coupling, the solid red, green and blue lines in both figure represent the constraints for the labeled fixed value of the other coupling.  The shaded area represents the new constraints. The three lines overlap near the upper right corner. The dashed lines on the left represent new constrains with pseudoscalar Yukawa coupling. The solid black lines in the figure on the left represents previous experimental constrains on $g _{ae}$ from XENONnT \cite{XENONnT_2020} and red giants cooling \cite{Red-giant_2020}. The solid black lines in the figure on the right represents current experimental constrains on $g _{a \gamma  }$ from CAST \cite{CAST_2017} and H1821+643 \cite{H1821+643_2021}. The solid orange line on the right figure represents constraints coming from misalignment mechanism with $\mathcal{N}\sim 10 ^{4}$ \cite{generic_2020, Luca_2020}.} 
    \label{con1}
\end{figure*}

The velocity suppression of the contribution to $g-2$ can be softened if we consider the dark matter velocity, $|v_a|\sim 10^{-3}$  \cite{Foster_2018}, which was ignored in previous formulae. This additional contribution arises from an integral of the form
\begin{equation}
    \bar{I}_{ij }(k)\equiv g_{ae}^{2}\int  \dfrac{d^{4}\Xi_{q}}{(2\pi)^{3}}\dfrac{q_{i }q_{j }}{(q\cdot k)^{2}} ~.
\end{equation}
In the above expression, the denominator can be approximated as $q_0^2k_0^2$, since we are neglecting the momentum of the electron. If we take this approximation and a velocity distribution of dark matter, $f(q)$, which we assume to be spherically symmetric \footnote{See \cite{Foster_2018} for more details about the speed distribution.}, the off-diagonal elements of $\bar{I}_{ij }(k)$ will vanish to leading order. The diagonal pieces, on the other hand, are non-zero and give 
\begin{equation}
    \bar{I}_{ij }(k)=g_{ae}^{2}\int  \dfrac{d^{4}\Xi_{q}}{(2\pi)^{3}}\dfrac{q_{i } q_{j}}{(q\cdot k)^{2}}\simeq {\frac{1}{3}} v _{a}^{2} R I _{A}g_{ij}~,
\end{equation}
where again $v _{a}\sim 10^{-3}$ \cite{Foster_2018} and there is no sum on repeated indices. This piece and all background dependent pieces containing a $d^4 \Xi  _q$ will contribute to the dark matter velocity dependent terms to $\mathcal{W}(k)$, and we get a constraint which is an order of magnitude stronger, ($\sqrt{v _{a}^{2}/\frac{|\vec{k}|^{2}}{E^{2}}} \sim 10 ) $,\begin{equation}\label{result1}
    \left\lvert {\frac{1}{6}}\kappa v _{a}^{2}+2\zeta\right\rvert \leq \frac{2\alpha}{\pi}  \times 10^{-12}~.
\end{equation}

The new constraints obtained in this work on $g_{ae}$ are shown in the left of Fig. \ref{con1} for fixed values of $g_{a\gamma }$. For smaller $g_{a\gamma }$, the contribution from $\kappa$ dominates and the $\zeta$ contribution can be neglected. For example, for $g_{a\gamma}=10^{-20}~{\rm GeV}^{-1}$, corresponding to the dashed blue line on the left of Fig. \ref{con1}, the $\kappa$ term dominates giving the same constraint found in previous work \cite{Jason_2023}. If $g _{a \gamma }\gtrsim 10 ^{-19}\, \mathrm{GeV}^{-1}$, the constraints on $g_{ae}$ will be stronger than the current experimental limit. For fixed $g_{ae}$, Eq. (\ref{result1}) gives new constraints to $g_{a\gamma }$. 

Next, we consider constraints on these couplings from theoretical considerations. Dark matter production, for example, will place constraints on the allowed values for these couplings. If dark matter is produced by some misalignment mechanism \footnote{We set $\mathcal{N} \sim 10 ^{4}$ in this work, where we have $g \equiv \frac{\alpha}{2 \pi} \frac{1}{f_\phi} \mathcal{N}$, see \cite{generic_2020} for more details.}, the parameters must fall below the yellow line in the right side of Fig. \ref{con1}. Since this is the most likely production mechanism for axion-like dark matter, points above the yellow line are less compelling. In light of this constraint, if $g _{ae}$ is between $10 ^{-17}$ and $10 ^{-19}$, this method gives the strongest constrains on $g _{a \gamma }$ for $m _{a}\lesssim 10 ^{-18}\, \mathrm{eV}$.

\subsection{Derivative Coupling}

Next we turn to the derivative coupling found in Eq. (\ref{original}). If we repeat the same analysis above with a derivative axion-fermion vertex, the constraints are similar but not identical to that found above
\begin{equation}\label{derivative}
    \left\lvert -\frac{\kappa }{8}\frac{|\vec{k}|^{2}}{m _{e}^{2}}+2\zeta\right\rvert \leq \frac{2\alpha}{\pi} \times 10^{-12}~,
\end{equation}
if we neglect the dark matter velocity. However, because of the small electrons momentum $|\vec{k}| \approx  10 ^{-4}$, the constraint is again enhanced if we consider velocity of the axion like particle, $v _{a}\sim 10 ^{-3}$. In this case, the constraints can be enhanced,
\begin{equation}\label{result2}
    \left\lvert {-\frac{1}{3}}\kappa v _{a}^{2}+2\zeta\right\rvert \leq \frac{2\alpha}{\pi}  \times 10^{-12}~.
\end{equation}

For constraints on $\Bar{g}_{ae}$, they are shown in the left of Fig. \ref{con1} with solid lines, while the constraints on $g_{a\gamma}$ is identical to the pseudoscalar case shown in the right of Fig. \ref{con1}.

\subsection{Two-Loop Contribution}
\begin{figure}
    \centering
    \includegraphics[width=5cm]{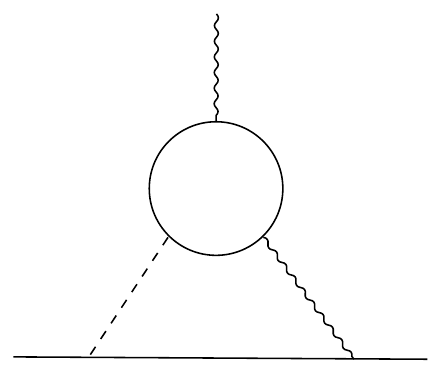}
    \hfill
    \caption{Two-loop Barr-Zee diagram contribute to electron $g-2$.}
    \label{two-loop}
\end{figure}

\begin{figure*}
    \centering
    
    \includegraphics[width=0.48\textwidth]{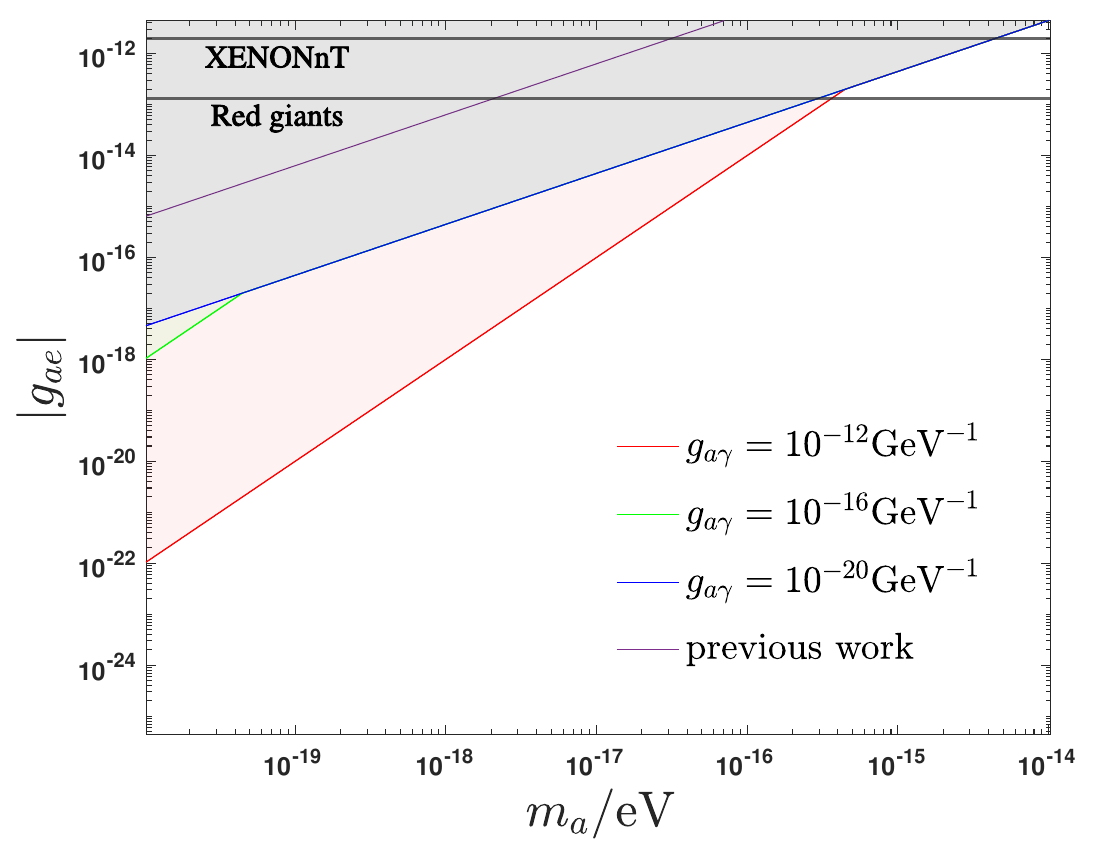}

    \caption{Constrains on the axion-electron pseudoscalar coupling from adding the two-loop contributions. The solid red, green and blue lines represent new constrains for different fixed values of $g_{a\gamma}$. The shaded region represents the forbidden parameter space. The solid black lines represents current experimental constrains on $g _{ae}$ from XENONnT \cite{XENONnT_2020} and red giants \cite{Red-giant_2020}. The purple line represents the previous result in \cite{Jason_2023}. }
    \label{con3}
\end{figure*}
Since the suppression factors of the one-loop contributions to $g-2$ that depend on $g_{ae}^2$ are much smaller than a loop factor, i.e. $\frac{|\vec{k}|^{2}}{E^{2}} \sim 10 ^{-8} $ and $v _{a}^{2}\sim 10^{-6}$, it is not unreasonable to expect that contributions from higher order loops will dominate. 

\begin{figure}
    \centering
    \includegraphics[width=10cm]{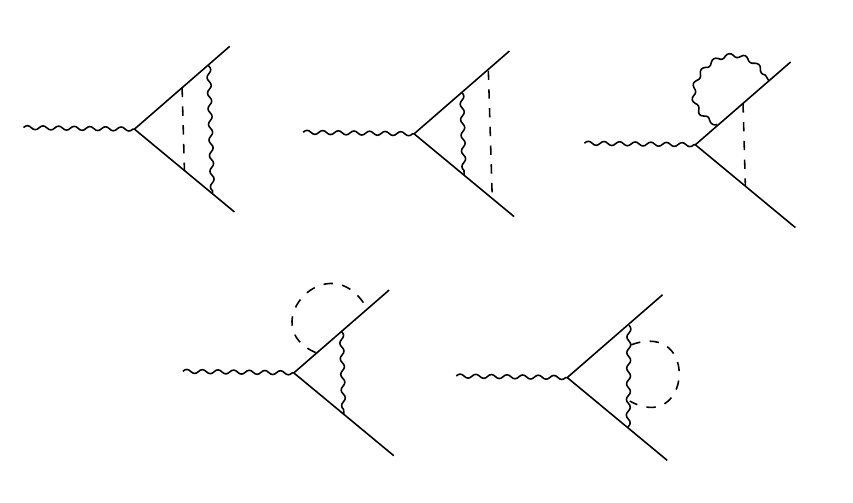}
    \hfill
    \caption{Other two-loop diagrams contribute to electron $g-2$.}
    \label{others}
\end{figure}

Here, we will briefly consider the contribution to $g_e-2$ at two-loops. We do not spend the time to calculate all two-loop contributions, but instead calculate the diagram found in Fig. \ref{two-loop} and use it as a way to estimate the size of all of the two-loop diagrams found in Fig. \ref{others}. We will leave the calculation of all these diagrams to future work. The only way this estimate for the two-loop contribution will break down if there is some kind of cancellation between the diagram in Fig. \ref{two-loop} and those in Fig. \ref{others}. However, this cancellation is extremely unlikely due to the fact that the fermion loop in Fig. \ref{two-loop} need not be a loop composed of electrons. In fact, it could be composed of muons or taus, even for flavor diagonal couplings of the axion-like particle. On the other hand, the diagrams in Fig. \ref{others} can only involve a muon through flavor non-universal axion like particle couplings. Thus, the diagrams in Fig. \ref{two-loop} and Fig. \ref{others} are truly independent and must renormalize independently. Thus, a cancellation is highly unlikely and our calculation should give a good order of magnitude estimate.

If we replace the $\kappa$ contribution with the contribution from the diagram in Fig. \ref{two-loop}, the constraint on the pseudoscalar coupling from the anomalous magnetic moment becomes 
\begin{equation}\label{result11}
    \left\lvert -\tilde{\kappa }+2\zeta\right\rvert \leq \frac{2\alpha}{\pi} \times 10^{-12}~,
\end{equation}
where 
\begin{equation}
    \tilde{\kappa }=\frac{\alpha}{\pi}\cdot \frac{g_{ae}^{2}\rho_{a}}{m_{a}^{2}m_{e}E}~. \label{eq:twoloopCon}
\end{equation}
As hoped, this contribution is not suppressed by the electron or dark matter velocities. This leads to a constraint which is stronger by a factor of $\sqrt{{\alpha}/v  _{a}^{2}} \sim 10 ^{2} $. 

The constraints on the derivative couplings are unchanged by including the two-loop contribution in Fig. \ref{two-loop}, since this contribution vanishes for a derivative coupling. This is due to the fact that the fermion loop breaks the axion shift symmetry. For a theory which only couples through the derivative coupling, this symmetry is exact and can't be broken \cite{Buen_2021}. Thus, this contribution must vanish.

The new constraints on the pseudoscalar coupling, when the two-loop contribution is included, are shown in Fig. \ref{con3}. We see that the constraints on $g _{ae}$ are enhanced by about $10 ^{2}$ compared to the small $g _{a \gamma }$ limit in Fig. \ref{con1}. The purple line in Fig. \ref{con3} represents the previous limit in \cite{Jason_2023}. Since the one-loop contribution to $g-2$ from axion-photon-photon coupling is not velocity suppressed, the two-loop contributions should be subdominate and can be ignored.  

\section{Conclusions}\label{5}
An important effect that has been neglected until recently \cite{Jason_2022,Jason_2023}, is the Bose enhancement of a propagating boson in a background of itself.  This Bose enhancement can elevate seemingly benign contributions to the level that they can be seen in precision measurements. This makes it possible to use precision measurements to search for ultralight dark matter.

Here, we continue this reasoning and discussed the axion dark matter background enhanced correction to the electron $g-2$. Full one-loop diagrams are calculated for contributions involving the axion-like particles couplings to photons and the electron. These contributions are compared to the uncertainties in the electron $g-2$ measurement. Due to the precise measurement of the electrons $g-2$ and the strong enhancement to these processes from the background, new strong constraints are realized on  $g _{ae}$ and $g _{a \gamma }$ as shown in Fig. \ref{con1} and Fig. \ref{con3}. 

Since the contribution to the electron $g-2$ involving the electron axion coupling is velocity suppressed at one-loop, we consider the effect of the two-loop diagrams. We do not do an exhaustive study at the two-loop level but instead calculate a subset of diagrams and argue why this should be a good estimate.  Since this contribution is two orders of magnitude larger then previous contributions, it should act as a good estimate of how large this contribution is and what kind of constraints to expect from $g-2$ experiments.

\appendix
\section{Derivation of Vertex Correction}\label{process}
For the first three diagrams in Fig. \ref{vertex1.pdf}, we get 
\begin{equation}
    \begin{aligned}
    & iZ_2^{-\frac{1}{2}}(\bar{k}) Z_2^{-\frac{1}{2}}(k) M_{\mu }^{\mathrm{Tree}} \\
    & =-i e \bar{u}(\bar{k}) \gamma_\mu u(k)\left[1+\frac{1}{2} C(k)+\frac{1}{2} \frac{m}{E} \frac{d}{d E}\left(\frac{k \cdot D(k)}{m}\right)-\frac{1}{2} \frac{D^0(k)}{E}\right. \\
    & \left.\quad+\frac{1}{2} C(\bar{k})+\frac{1}{2} \frac{m}{E} \frac{d}{d E}\left(\frac{k \cdot D(\bar{k})}{m}\right)-\frac{1}{2} \frac{D^0(\bar{k})}{E}\right]~,
    \end{aligned}
\end{equation}
the self-energy contribution takes the form
\begin{equation}
    \begin{aligned}
        i\mathcal{M}_{\mu }^{\mathrm{SE}}&=i\Sigma(\bar{k})\cdot \dfrac{i}{\slashed{\bar{k}}-m_{e} }\cdot(-ie)\gamma_{\mu }+(-ie)\gamma_{\mu }\cdot \dfrac{i}{\slashed{k}-m_{e} } \cdot i\Sigma(k)\\    
        &=ie \left[\eval{\left(C+\dfrac{m}{E}\dfrac{d}{dE}\left(\dfrac{p\cdot D}{m}\right)-\dfrac{D^{0}}{E}\right)}_{\slashed{\bar{k}}=m}+\eval{\slashed{D}(\bar{k})}_{\slashed{\bar{k}}=m}\dfrac{1}{\bar{\slashed{k}}-m}\right]\gamma_{\mu }+\text{conjugate part}~,
    \end{aligned}~.
\end{equation}
with the counter term giving
\begin{equation}
    i\mathcal{M}_{ \mu }^{\mathrm{CT}}=-i \eval{\slashed{D}(\bar{k})}_{\slashed{\bar{k}}=m} \cdot\dfrac{i}{\slashed{\bar{k}}-m_{e}}\cdot(-ie)\gamma_{\mu }+(-ie)\gamma_{\mu }\cdot \dfrac{i}{\slashed{k}-m_{e} } \cdot \left(-i \eval{\slashed{D}(k)}_{\slashed{k}=m}\right)~,
\end{equation}
where we expanded the self-energy about the pole mass. The total contribution then reads
\begin{equation}
    \begin{aligned}
        i\mathcal{M}^{\mathrm{\uppercase\expandafter{\romannumeral1}}}_{\mu }&\equiv  i\mathcal{M}^{\mathrm{Tree}}_{\mu }+i\mathcal{M}^{\mathrm{SE}}_{\mu }+i\mathcal{M}^{\mathrm{CT}}_{\mu }\\   
        &=(-ie)\bar{u}_{n}(\bar{k}) \gamma_{\mu }\left[1-\frac{1}{2}C(k)-\frac{1}{2}\dfrac{m}{E}\dfrac{d}{dE}\left(\dfrac{k\cdot D(k)}{m}\right)+\frac{1}{2}\dfrac{D_{0}(k)}{E}+\left(k\leftrightarrow \bar{k}\right)\right]u_{n}(k)~,
    \end{aligned}
\end{equation}
this is what we obtained in Eq. (\ref{123}). 

Now we consider diagrams 4 and 5 in Fig. \ref{vertex1.pdf}. Diagram 4 gives 
\begin{equation}
    i\mathcal{M}^{4}_{\mu }=-i e\left[\frac{1}{2} \bar{u}(\bar{k})\left[\frac{d \Sigma_n(k)}{d k_\mu}+\frac{d \Sigma_n(\bar{k})}{d \bar{k}_\mu}\right] u(k)+\bar{u}(\bar{k}) F _{\mu }^{A} \left(k,\Delta k\right) u(k)\right]~.
\end{equation} 
Using Gordon decomposition on the wave function renormalization, we have
\begin{equation}
    \begin{aligned}
        \frac{d \Sigma_n(k)}{d k_\mu}&=\gamma_\mu C(k)+\gamma_\nu \frac{d D^\nu(k)}{d k_\mu}\\    
        &= \gamma_\mu C(k)+\frac{d }{d k_\mu}\left(\dfrac{k\cdot D(k)}{m_{e}}\right)-\dfrac{D_{\mu }(k)}{m_{e}}+F _{\mu }^{B}\left(k,\Delta k\right)~,\\  
    \end{aligned}
\end{equation}
where
\begin{equation}
    \begin{aligned}
        F _{\mu }^{A}\left(k,\Delta k\right)&= \dfrac{\Delta k^{\nu }}{2m_{e}}\left[i\sigma_{\mu\nu }\cdot \dfrac{1}{2}R \slashed{\bar{I}}(k)+\dfrac{1}{2}R\bar{I}_{\nu }(k)\right]  \\
        F _{\mu }^{B}\left(k,\Delta k\right)&=\dfrac{i\sigma_{\mu\nu }(\Delta k)^{\nu }}{2m_{e}}\frac{d D^\nu(k)}{d k_\mu}~.
    \end{aligned}
\end{equation}
Diagram 5 of Fig. \ref{vertex1.pdf} gives
\begin{equation}
    i\mathcal{M}_{\mu }^{5}=(-ie)\bar{u}(\bar{k})F _{\mu }^{C}\left(k,\Delta k\right)u(k)~,
\end{equation}
where
\begin{equation}
    F _{\mu }^{C}\left(k,\Delta k\right)=\left[-2I^{\mu }_{\nu}+2 R I_{B}\delta^{\mu }_{\nu }\right]\times \dfrac{i \sigma^{\nu \alpha}(\Delta k)_{\alpha} }{2m_{e}}~.
\end{equation}
Combining the $F_\mu(\Delta k)$ corrections of each vertex we get
\begin{equation}
    \begin{aligned}
        F_{\mu }\left(k,\Delta k\right)& \equiv  F _{\mu }^{A}\left(k,\Delta k\right)+ F _{\mu }^{B}\left(k,\Delta k\right) +F _{\mu }^{C}\left(k,\Delta k\right)\\
        &=\dfrac{\Delta k^{\alpha }}{2m_{e}}\left[i\sigma_{\nu  \alpha }\cdot \left(\dfrac{1}{2}R \slashed{\bar{I}}(k)\delta^{\nu }_{\mu }-\frac{1}{4}\bar{I}_\mu^\nu+\frac{1}{2}R \frac{k_\mu}{m_e} \bar{I}^\nu(k)-2I^{\nu }_{\mu}+2 R I_{B}\delta^{\nu }_{\mu }\right)+\dfrac{1}{2}R\bar{I}_{\alpha }(k)\right]~.\\
    \end{aligned}
\end{equation}
The total vertex correction is then
\begin{equation}
    \begin{aligned}
        i\mathcal{M}^{\mathrm{TOT}}_{\mu }   
        &=(-ie)\bar{u}_{n}(\bar{k})\left[\gamma_{\mu }+G_{\mu }(k)+G_{\mu }(\bar{k})+F_{\mu }\left(k,\Delta k\right)\right]u_{n}(k)~,
    \end{aligned}
\end{equation}
where 
\begin{equation}
    G_{\mu }(k)\equiv  \frac{1}{2}\left(\dfrac{d}{d k^\mu}-\gamma_{\mu }\dfrac{m}{E} \frac{d}{d E}\right)\left(\frac{k \cdot D(k)}{m_e}\right)+\gamma_{\mu }\frac{D_{0}(k)}{2E}-\frac{D_{\mu }(k)}{2m_{e}}~.
\end{equation}
This last piece contributes to the anomalous magnetic moment with a contribution 
\begin{eqnarray}
    G_{\mu}(k)\supset \dfrac{\Delta k^{\alpha }}{8m_{e}}i\sigma_{\nu  \alpha }R I_{A}\delta^{\nu }_{\mu }~,
\end{eqnarray}
and all other contribution are found in $F_{\mu }\left(k,\Delta k\right)$.

\section{Derivation of the Hamiltonian}\label{Ham}
To determine the Hamiltonian, first we define 
\begin{equation}
    \rho_1=-\gamma_5 \qquad  \rho_2=i \gamma_0 \gamma_5 \qquad \rho_3=\gamma_0
\end{equation}
The modified Dirac equation can now be written in terms of these matrices as follow
\begin{equation}
    \begin{aligned}
    & {\left[\slashed{k}+\gamma_0 D^0-m_e\right.} 
     -e \vec{\gamma} \cdot \vec{A}\left[1+\frac{D^0(k)}{E}\right] \\
    & +\frac{e}{2m_{e}}\left(\partial^j A^i\right)\left[ R \frac{k_i}{2m_e} \bar{I}^0(k)i\rho_{1} \sigma_{j}-\frac{1}{2} R \rho_{3} \bar{I}^0(k) \sigma_{ij}\right. \\
    & \left.\left.+\left(\frac{1}{2} R  I_A(k)-2R I_{B}(k)+\frac{1}{4}\bar{I}_{ii}(k)\right)\sigma_{ij}\right] \right] u_n(k)=0~.
    \end{aligned}
\end{equation}
Here we choose the magnetic field $\vec{A}=\frac{1}{2}\vec{B}\times \vec{r}$ to get the effective Hamiltonian
\begin{equation}\label{origin}
    H=-\rho_{1}\vec{\sigma}\cdot \vec{\pi}-D^{0}+\rho_{3}m_e-\frac{e}{2m_{e}}\left[ \mathcal{A}_{1} + \mathcal{A}_{2}+ \mathcal{A}_{3} \right] ~,
\end{equation}
where
\begin{equation}
    \begin{aligned}
        \vec{\pi}&=\vec{k}-e\vec{A}\left(1+\frac{D^0(k)}{E}\right)\\
        \mathcal{A}_{1}&=  \frac{1}{4m_{e}}R\bar{I}^0(k)\left(\vec{\sigma} \times \vec{B}\right) \cdot \vec{k} \rho_2          \\  
        \mathcal{A}_{2}&=  \frac{1}{2}R\bar{I}^{0}\left(\vec{\sigma}\cdot \vec{B}\right)           \\  
        \mathcal{A}_{3}&=  \left(2RI_{B}(k)-\frac{1}{2}R I_A(k)-\frac{1}{4}\bar{I}_{ii}(k)\right)\left(\vec{\sigma}\cdot \vec{B}\right) \rho_{3}  ~, \\
    \end{aligned}
\end{equation}
after the Foldy-Wouthuysen transformation \cite{Donoghue_1985}, we have
\begin{equation}
    \mathcal{H}=E-\frac{e}{2E}\left(\vec{L}\cdot \vec{B}\right)\left(1+\frac{D^0(k)}{E}\right)-\frac{e}{2E}\left(\vec{\sigma}\cdot \vec{B}\right)\left(1+\frac{D^0(k)}{E_k}+\mathcal{W}(k) \right)+\text{off diagonal}~.
\end{equation}
The off-diagonal pieces can be ignored since they are higher order after the transformation. Strictly speaking, there are additional pieces that contribute beyond those in Eq. (\ref{effH}). These other pieces are proportional to $(\vec{\sigma }\cdot \hat{k})(\hat{k}\cdot \vec{B})$, which are relevant when $\vec{k}$ is parallel to $\vec{B}$. We can ignore these pieces, since the experiment is designed such that the dominant velocity is perpendicular to the magnetic field.

\bibliography{biblio.bib}
\providecommand\begingroup\raggedright
\end{document}